%% file: DESY-06-125.tex
\documentclass[zpreprint,zbstdefault]{zeus_paper}
\usepackage[english]{babel}
\input ./zeus_def.tex
\input  DESY-06-125-cit.tex
\begin{document}
\include{DESY-06-125-tit}

\include{auth133_out}

\include{DESY-06-125-txt}
\include{DESY-06-125-ref}
%
\include{DESY-06-125-tab}
%
\include{DESY-06-125-fig}

%
\end{document}

%% file: zeus_def.tex
\chardef\usc=95
\chardef\til=126
\catcode`\@=11 
\DeclareRobustCommand\xdotspace{\futurelet\@let@token\@xdotspace}
\def\@xdotspace{%
  \ifx\@let@token.\else
  \ifx\@let@token\bgroup.\else
  \ifx\@let@token\egroup.\else
  \ifx\@let@token\/.\else
  \ifx\@let@token\ .\else
  \ifx\@let@token~.\else
  \ifx\@let@token!.\else
  \ifx\@let@token,.\else
  \ifx\@let@token:.\else
  \ifx\@let@token;.\else
  \ifx\@let@token?.\else
  \ifx\@let@token/.\else
  \ifx\@let@token'.\else
  \ifx\@let@token).\else
  \ifx\@let@token-.\else
  \ifx\@let@token\@xobeysp.\else
  \ifx\@let@token\space.\else
  \ifx\@let@token\@sptoken.\else
   .\space
   \fi\fi\fi\fi\fi\fi\fi\fi\fi\fi\fi\fi\fi\fi\fi\fi\fi\fi}
\catcode`\@=12 

\newcommand{\stru}[2]{%
   \relax\ifmmode\hbox{\vrule height#1 depth#2 width0pt}%
   \else\vrule height#1 depth#2 width0pt\fi}

\newcommand{\Ronum}[1]{\uppercase\expandafter{\romannumeral#1}}
\newcommand{\ronum}[1]{\expandafter{\romannumeral#1}}
\DeclareRobustCommand{\LaTeXZ}{%
  \LaTeX\kern-.05em4\kern-.1em
  {\raisebox{-0.2ex}{$\scriptstyle\text{ZEUS}$}}\xspace}

\DeclareMathAlphabet{\mathbf}{OT1}{cmr}{bx}{sl}
\newcommand{\eVdist}{\kern-0.06667em}

\newcommand{\gev}{{\,\text{Ge}\eVdist\text{V\/}}}

\newcommand{\pb}{\,\text{pb}}

\newcommand{\slashfrac}[2]{%
  \raisebox{0.5ex}{\ensuremath #1}\kern-0.12em/\kern-0.08em
  \raisebox{-.8ex}{\ensuremath #2}}

\newcommand{\sqr}[3]{%
    {\vcenter{\hrule height.#3ex\hbox{\vrule width.#2ex height#1ex
     \kern#1ex\vrule width.#3ex}\hrule height.#2ex}}}

\catcode`\@=11 
\newcommand{\parenbar}{\mathpalette\p@renb@r}
\def\p@renb@r#1#2{\vbox{%
  \ifx#1\scriptscriptstyle \dimen@.7em\dimen@ii.2em\else
  \ifx#1\scriptstyle \dimen@.8em\dimen@ii.25em\else
  \dimen@1em\dimen@ii.4em\fi\fi \offinterlineskip
  \ialign{\hfill##\hfill\cr
    \vbox{\hrule width\dimen@ii}\cr
    \noalign{\vskip-.3ex}%
    \hbox to\dimen@{$\mathchar300\hfil\mathchar301$}\cr
    \noalign{\vskip-.3ex}%
    $#1#2$\cr}}}
\catcode`\@=12 

\newcommand{\IP}{{\rm I$\kern-0.01667em$P}\xspace}

\mathchardef\qsm=63
\mathchardef\pls=43
\mathchardef\mns=512
\mathchardef\plm=518
\mathchardef\eql=61
\mathchardef\smallleft=300
\mathchardef\smallright=301
\mathchardef\les=316
\mathchardef\gre=318
\mathchardef\leq=532
\mathchardef\grq=533

\catcode`\@=11 
\newcounter{pict@width}
\newcounter{pict@height}
\newlength{\pict@scale}
\newcommand{\psfigadd}[4]{%
\setcounter{pict@width}{1*\ratio{#2+\pict@scale/2}{\pict@scale}}
\setcounter{pict@height}{1*\ratio{#3+\pict@scale/2}{\pict@scale}}
\setlength{\unitlength}{\pict@scale}
\hbox to #2{\hspace{-\fill}\begin{picture}(\thepict@width,\thepict@height)
\put(0,0){\psfig{figure=#1,width=#2,height=#3,clip=}}
\SetScale{0.283466457}
\SetWidth{1.763889}
{#4}
\end{picture}}
}
\newcounter{pict@widthfst}
\newcounter{pict@widthscd}
\newcounter{pict@widthtot}
\newcommand{\psfigaddtwo}[7]{%
\setcounter{pict@widthfst}{1*\ratio{#2+\pict@scale/2}{\pict@scale}}
\setcounter{pict@widthscd}{1*\ratio{#2+#4+\pict@scale/2}{\pict@scale}}
\setcounter{pict@widthtot}{1*\ratio{#2+#4+#6+\pict@scale/2}{\pict@scale}}
\setcounter{pict@height}{1*\ratio{#3+\pict@scale/2}{\pict@scale}}
\setlength{\unitlength}{\pict@scale}
\hbox{\hspace{-\fill}\begin{picture}(\thepict@widthtot,\thepict@height)
\put(0,0){\psfig{figure=#1,width=#2,height=#3,clip=}}
\put(\thepict@widthscd,0){\psfig{figure=#5,width=#6,height=#3,clip=}}
\SetScale{0.283466457}
\SetWidth{1.763889}
{#7}
\end{picture}}
}
\newcommand{\psfigror}[4]{%
\setcounter{pict@width}{1*\ratio{#2+\pict@scale/2}{\pict@scale}}
\setcounter{pict@height}{1*\ratio{#3+\pict@scale/2}{\pict@scale}}
\setlength{\unitlength}{\pict@scale}
\hbox{\begin{picture}(\thepict@width,\thepict@height)
\put(0,\thepict@height){\psfig{figure=#1,width=#3,height=#2,clip=,angle=270}}
\SetScale{0.283466457}
\SetWidth{1.763889}
{#4}
\end{picture}}
}
\newcommand{\psfigrol}[4]{%
\setcounter{pict@width}{1*\ratio{#2+\pict@scale/2}{\pict@scale}}
\setcounter{pict@height}{1*\ratio{#3+\pict@scale/2}{\pict@scale}}
\setlength{\unitlength}{\pict@scale}
\hbox{\begin{picture}(\thepict@width,\thepict@height)
\put(0,0){\psfig{figure=#1,width=#3,height=#2,clip=,angle=90}}
\SetScale{0.283466457}
\SetWidth{1.763889}
{#4}
\end{picture}}
}
\catcode`\@=12 
\newlength\listtextwidth

\catcode`\@=11 
\newlength{\@tabfninsert}
\newlength{\@tabfnwidth}
\newcommand{\tabfootnote}[2]{%
  \setlength{\@tabfninsert}{0.8em}
  \setlength{\@tabfnwidth}{\textwidth}
  \addtolength{\@tabfnwidth}{-\@tabfninsert}
  \addtolength{\@tabfnwidth}{-0.4em}
  \noindent\makebox[\@tabfninsert][r]{\footnotesize$^{#1}$\hfil}\hfill%
  \parbox[t]{\@tabfnwidth}{\footnotesize #2\hfill}}
\catcode`\@=12 

%% file: DESY-06-125-cit.tex
\def\citeCTD{{\cite{%
nim:a279:290,*npps:b32:181,*nim:a338:254%
}}\xspace}
\def\citeCAL{{\cite{%
nim:a309:77,*nim:a309:101,*nim:a321:356,*nim:a336:23%
}}\xspace}

%% file: DESY-06-125-tit.tex
\prepnum{DESY 06-125}

\title{
Measurement of prompt photons \\ with associated
jets \\ in  photoproduction at HERA
}

\author{ZEUS Collaboration}
\date{September 14, 2006}

\abstract{
The photoproduction of prompt photons, together with an accompanying
jet, has been studied in $ep$ collisions at a centre-of-mass energy
of 318 GeV with the ZEUS detector at HERA using an integrated
luminosity of 77 pb$^{-1}$.
Cross sections were measured 
for the transverse energy of the photon and the jet larger than 5 and 6 $\gev$, respectively.
The differential $\gamma$+jet cross sections  
were reconstructed as  
functions of the transverse energy,  pseudorapidity
and $x_{\gamma}^{\rm obs}$, the fraction of the incoming photon 
momentum taken by the photon-jet system.
Predictions based on leading-logarithm
parton-shower Monte Carlo models  and next-to-leading-order (NLO) QCD
generally underestimate the cross sections for the transverse energies
of prompt photons below $7\gev$, while the $k_T$-factorisation QCD calculation
agrees with the data better. When the minimum transverse energy of
prompt photons is increased to $7\gev$, both NLO QCD and the $k_T$-factorisation calculations 
are in good agreement with the data.
}
\makezeustitle

%% file: auth133_out.tex
%
%
%
%
\pagenumbering{Roman}
\begin{center}                                                                                     
{                      \Large  The ZEUS Collaboration              }                               
\end{center}                                                                                       
  S.~Chekanov,                                                                                     
  M.~Derrick,                                                                                      
  S.~Magill,                                                                                       
  S.~Miglioranzi$^{   1}$,                                                                         
  B.~Musgrave,                                                                                     
  D.~Nicholass$^{   1}$,                                                                           
  \mbox{J.~Repond},                                                                                
  R.~Yoshida\\                                                                                     
 {\it Argonne National Laboratory, Argonne, Illinois 60439-4815}, USA~$^{n}$                       
\par \filbreak                                                                                     
  M.C.K.~Mattingly \\                                                                              
 {\it Andrews University, Berrien Springs, Michigan 49104-0380}, USA                               
\par \filbreak                                                                                     
  N.~Pavel~$^{\dagger}$, A.G.~Yag\"ues Molina \\                                                   
  {\it Institut f\"ur Physik der Humboldt-Universit\"at zu Berlin,                                 
           Berlin, Germany}
% ...
\newpage                                                                                           
$^{\    1}$ also affiliated with University College London, UK \\                                  
% ...
$^{\dagger}$ deceased \\                                                                           
%
\newpage   
                                                           %
                                                           %
\begin{tabular}[h]{rp{14cm}}                                                                       
$^{a}$ &  supported by the Natural Sciences and Engineering Research Council of Canada (NSERC) \\  
$^{b}$ &  supported by the German Federal Ministry for Education and Research (BMBF), under        
          contract numbers HZ1GUA 2, HZ1GUB 0, HZ1PDA 5, HZ1VFA 5\\                                
$^{c}$ &  supported in part by the MINERVA Gesellschaft f\"ur Forschung GmbH, the Israel Science   
          Foundation (grant no. 293/02-11.2) and the U.S.-Israel Binational Science Foundation \\  
$^{d}$ &  supported by the German-Israeli Foundation and the Israel Science Foundation\\           
$^{e}$ &  supported by the Italian National Institute for Nuclear Physics (INFN) \\                
$^{f}$ &  supported by the Japanese Ministry of Education, Culture, Sports, Science and Technology 
          (MEXT) and its grants for Scientific Research\\                                          
$^{g}$ &  supported by the Korean Ministry of Education and Korea Science and Engineering          
          Foundation\\                                                                             
$^{h}$ &  supported by the Netherlands Foundation for Research on Matter (FOM)\\                   
$^{i}$ &  supported by the Polish State Committee for Scientific Research, grant no.               
          620/E-77/SPB/DESY/P-03/DZ 117/2003-2005 and grant no. 1P03B07427/2004-2006\\             
$^{j}$ &  partially supported by the German Federal Ministry for Education and Research (BMBF)\\   
$^{k}$ &  supported by RF Presidential grant N 1685.2003.2 for the leading scientific schools and  
          by the Russian Ministry of Education and Science through its grant for Scientific        
          Research on High Energy Physics\\                                                        
$^{l}$ &  supported by the Spanish Ministry of Education and Science through funds provided by     
          CICYT\\                                                                                  
$^{m}$ &  supported by the Particle Physics and Astronomy Research Council, UK\\                   
$^{n}$ &  supported by the US Department of Energy\\                                               
$^{o}$ &  supported by the US National Science Foundation\\                                        
$^{p}$ &  supported by the Polish Ministry of Scientific Research and Information Technology,      
          grant no. 112/E-356/SPUB/DESY/P-03/DZ 116/2003-2005 and 1 P03B 065 27\\                  
$^{q}$ &  supported by FNRS and its associated funds (IISN and FRIA) and by an Inter-University    
          Attraction Poles Programme subsidised by the Belgian Federal Science Policy Office\\     
$^{r}$ &  supported by the Malaysian Ministry of Science, Technology and                           
Innovation/Akademi Sains Malaysia grant SAGA 66-02-03-0048\\                                       
\end{tabular}                                                                                      
                                                           %
                                                           %

%% file: DESY-06-125-txt.tex
\pagenumbering{arabic}
\pagestyle{plain}
\section{Introduction}
\label{sec-int}

Events containing an isolated photon (prompt photon) are a powerful tool
to study hard interaction processes since such photons
emerge without the hadronisation phase by which a final state quark or
gluon forms a jet.
In $ep$ collisions, the presence of a jet in addition to the photon
allows  measurements that are more sensitive to the underlying
partonic processes than is possible for inclusive prompt-photon events.
In particular, final states with a prompt photon with
a high transverse energy ($E_{T}$)
together with a high-$E_{T}$ jet are directly sensitive
to the quark content of the proton through the scattering of the exchanged photon
with a quark, $\gamma q\to\gamma q$ (Compton scattering).
In this case, the incident
photon is point like, and the process (Fig.~\ref{prph_fe} (a-b)) is called direct.
For exchanged four-momentum transfer
close to zero (photoproduction),
the additional contribution to prompt-photon
events from the $gq \to q\gamma$ process,
in which one of the
initial partons comes from the photon which displays a hadronic
structure (resolved process, Figs.~\ref{prph_fe} (c-d)),
can be dominant
\cite{pr:d52:58, ejp:c21:303, *ejp:c34:191, pr:d64:14017,*gamjet}.
Prompt-photon measurements can be used to constrain the parton
distribution functions (PDFs) of the proton and of the photon,
as well as provide a testing ground for QCD calculations.
A number of predictions exist
\cite{pr:d52:58, ejp:c21:303, *ejp:c34:191,  pr:d64:14017, *gamjet, Lipatov:2005tz} that
can be confronted  with the data.

The first observation by ZEUS of isolated photons accompanied by a hadronic jet in
photoproduction used an integrated
luminosity of 6.4~$\mathrm{\mathrm{pb^{-1}}}$\cite{pl:b413:201}.
Distributions sensitive to the intrinsic $k_{T}$ in the
$\gamma$+jet final state were later
measured by ZEUS~\cite{pl:b511:19}.
Inclusive prompt-photon cross sections
with  no jet requirement have also been reported \cite{pl:b472:175}.
Recently, H1 have published results on the $\gamma$+jet final state
in photoproduction~\cite{H1prompt}.

This paper reports the first ZEUS results on
differential cross sections of the $\gamma$+jet final state in
the photoproduction regime of $ep$ scattering.
The cross sections are presented as a function of
the transverse energy and pseudorapidity of both the photon
($E_T^{\gamma}$, $\eta^{\gamma}$) and
the jet ($E_T^{\rm jet}$, $\eta^{\rm jet}$), as well as the
fraction of the incoming photon momentum taken by the photon-jet system
($x_{\gamma}^{\rm{obs}}$).
In contrast to  previous
measurements~\cite{pl:b413:201,pl:b511:19,pl:b472:175,H1prompt},
the present analysis is based
on the conversion-probability method, which uses information on the frequency
with which photons convert to $e^+e^-$ in front of a
dedicated preshower detector.
Cross sections for $\gamma$+jet events are  compared
to next-to-leading-order (NLO) QCD, 
calculations based on
the $k_T$-factorisation approach and Monte Carlo (MC) models incorporating
leading-order matrix elements plus parton showers.
Since jets at relatively low  $E_T^{\rm jet}$ are measured in addition to the photon,
parton-level calculations were corrected for hadronisation effects using a MC model.
The hadronisation  correction for $\gamma$+jet events is expected to be smaller
than for dijets with similar jet transverse energies since
the photon does not undergo  hadronisation. Therefore, for low $E_T^{\rm jet}$,  
the theoretical predictions for the $\gamma$+jet cross sections
are expected to be more reliable than for dijet final states.

\section{Data sample and experimental setup}

The data sample was  taken
during the 1999-2000 period, corresponding to an integrated luminosity of
$77.1\pm 1.6$  pb$^{-1}$.
The positron or electron beam energy was $27.5$ GeV and the proton beam
energy was $920$ GeV, corresponding to a centre-of-mass energy of 318 GeV.
Here and in the following, the term ``electron'' denotes generically both the electron ($e ^-$)
and the positron ($e^+$), unless otherwise stated.

ZEUS is a multipurpose detector described in detail
elsewhere \cite{zeus:1993:bluebook}.
Of particular importance in the
present study  are the central tracking detector,
the uranium-scintillator calorimeter and the barrel preshower detector.

The central tracking detector (CTD) \citeCTD is  a cylindrical
drift chamber with nine super-layers covering the
polar-angle\footnote{
The ZEUS coordinate system is a right-handed Cartesian system, with
the $Z$ axis pointing in the proton beam direction, referred to
as the ``forward direction'', and the $X$ axis
pointing left towards the centre of HERA. The coordinate origin is at the
nominal interaction point.}
region $15^\circ < \theta < 164^\circ$ and the
radial range $18.2 - 79.4$ cm. Each super-layer consists
of eight sense-wire layers. The transverse-momentum resolution
for charged tracks traversing all CTD layers is
$\sigma(p_{T})/p_{T} = 0.0058 p_{T}
\oplus 0.0065 \oplus  0.0014/p_{T}$,
with $p_{T}$ in GeV.

The CTD is  surrounded by the uranium-scintillator
calorimeter, CAL \citeCAL, which is divided
into three parts: forward, barrel and rear with the barrel consisting of 32 modules.
The calorimeter is longitudinally segmented into electromagnetic
(EMC) and hadronic (HAC) sections.
The smallest subdivision of the CAL is called a
cell. The energy resolution of the calorimeter
under test-beam conditions
is $\sigma_E/E=0.18/\sqrt{E}$ for electrons and
$\sigma_E/E=0.35/\sqrt{E}$ for hadrons, with $E$ in GeV.

The luminosity was measured using the bremsstrahlung process $ep \to e
p \gamma$ with the luminosity
monitor~\cite{Desy-92-066,*zfp:c63:391,*acpp:b32:2025}, a
lead--scintillator calorimeter placed in the HERA tunnel at $Z = -107$ m.

The ZEUS barrel preshower detector (BPRE) \cite{magill:bpre}
is located in front of the barrel calorimeter.
The BPRE detector consists of 32 cassettes each containing
13 scintillator tiles of size $20\times 20$ cm that were
installed directly
in front of each of the 32 barrel CAL modules.
The measured output, calibrated in minimum ionising particle units (mips), is proportional
to the energy loss of the incident particle after interaction with  material
(mainly the superconducting coil)  in front of the barrel calorimeter.

The mip calibration of each of the $416$ channels of the BPRE was done using
all events triggered in the ZEUS detector.
A luminosity of approximately 1 $\pb^{-1}$ was required for each calibration run.
The one-mip signal was validated using  cosmic-ray muon data.
The single-mip resolution was measured to be $0.3$ mips 
and a minimum charge threshold corresponding to this value was applied
to each channel.
After calibration and correction for dead or inefficient channels,
the signal efficiency for  scattered electrons from
deep-inelastic events was larger than 99$\%$.

\section{Theoretical predictions
\label{sec:NLO}}

The measured $\gamma$+jet cross sections  were  compared
to NLO QCD based on collinear factorisation and DGLAP
evolution~\cite{sovjnp:15:438,*sovjnp:20:94,*jetp:46:641,*np:b126:298}, as well as
to calculations based on
the $k_T$-factorisation approach with unintegrated quark and gluon densities.

A NLO calculation with additional higher-order terms  was performed
by Krawczyk and Zembrzuski (KZ) \cite{pr:d64:14017,*gamjet}.
The calculation includes the leading-order term $\gamma q
\rightarrow \gamma q$, $\alpha_S$ corrections to this term, initial and
final resolved-photon contributions, double-resolved contributions
and the direct box diagram $\gamma g \rightarrow \gamma g$.
The latter two
contributions are calculated to order $\alpha_s^2$.
No intrinsic transverse momentum of the initial-state
partons in the proton was assumed.
The renormalisation and factorisation scales for such calculation
are set to $\mu_{R}=\mu_{F} = E_{T}^{\gamma}$.
The GRV parameterisation of the proton PDF  \cite{zfp:c67:433},
the photon PDF  \cite{pr:d45:3986,*pr:d46:1973} and the
fragmentation function \cite{pr:d48:116}, were used.

A similar NLO calculation
by Fontannaz, Guillet and Heinrich (FGH)~\cite{ejp:c21:303, *ejp:c34:191}
contains additional higher-order corrections
to the resolved photon process.
For the FGH calculation, the MRST01 \cite{epj:c14:133, *Martin:2001es} proton PDF
and the AFG02 \cite{zfp:c64:621} photon PDF were used.

The prediction of A.~Lipatov and N.~Zotov (LZ)~\cite{Lipatov:2005tz}
is  based on the
$k_T$-factorisation~\cite{sovjnp:53:657,*np:b366:135,*np:b360:3} method.
The LZ calculation  uses  the  unintegrated quark and gluon
densities of the  proton and photon according to the
Kimber-Martin-Ryskin (KMR) prescription \cite{Kimber:2001sc,*Watt:2003mx}
with the GRV parametrisations \cite{zfp:c67:433,pr:d45:3986,*pr:d46:1973} of
collinear quark and gluon densities.
In this approach, both direct and resolved contributions
are taken into account.

For all the calculations discussed, jets were
reconstructed by running the longitudinally invariant $k_{T}$
cluster algorithm in the inclusive mode \cite{pr:d48:3160,*np:406:187} on partons.
A prompt-photon jet was defined as a jet containing the  final-state photon. An isolation requirement,
$E_{T}^{\gamma, \rm (true)}>0.9\, E_{T}^{\gamma}$, where $E_{T}^{\gamma, \rm (true)}$  is the
transverse energy of the final-state photon and  $E_{T}^{\gamma}$  is the
total transverse energy
of the prompt-photon jet, was applied to avoid the effects of collinear
photon emission from quarks and to match the analysis isolation requirement
(see Section~\ref{data}). A comparison with NLO calculations based on a cone
isolation requirement showed consistent results \cite{Andzei}.

The calculations were  corrected for hadronisation
effects using  the {\sc Pythia}  MC model discussed
in Section~\ref{sec:Monte-Carlo-simulations}.
These corrections, which are negligible
in the case of inclusive prompt photons, cannot be neglected
when an accompanying jet is required.
The hadronisation correction factors were
defined as
$C^{\rm had}=\sigma(\mathrm{hadrons}) / \sigma(\mathrm{partons})$,
where $\sigma$ denotes the differential cross sections
calculated at the hadron and parton levels of the MC model, respectively.
For both the parton and hadron levels of the MC generated events,
the prompt photon was defined as the $k_T$ jet with at least one photon
and with the isolation requirement $E_{T}^{\gamma, \rm (true)}>0.9\, E_{T}^{\gamma}$.
The $E_{T}$ and $\eta$ distributions at the parton level in the MC model
have a different shape than in the NLO
calculations, especially at low  $E_{T}^{\gamma}$, where the NLO predictions rise faster
than do those of the MC.  To determine the hadronisation corrections,
the MC parton distributions were reweighted to match
the shapes of the NLO calculations. The reweighting was performed in four dimensions 
defined by the  $E_T^{\gamma}$, $\eta^{\gamma}$,  $E_T^{\mathrm{jet}}$ and $\eta^{\mathrm{jet}}$
variables.

The final  hadronisation correction was determined
from {\sc Pythia}  after the parton-level reweighting
procedure discussed above.
The {\sc Herwig} model discussed in Section~\ref{sec:Monte-Carlo-simulations}
requires a large reweighting so it was not used for the hadronisation correction.

The hadronisation correction factor for the total cross section in 
the kinematic range defined in
Section~\ref{result} was 0.92.
The corrections for the $\gamma+\rm jet$ differential cross sections
are close to unity  for large transverse momenta of the photon,
but they decrease to $0.78$ at low $E_T^{\gamma}$.
It was verified that
if no jet was required, the hadronisation corrections were close to unity.

The {\sc Pythia} default setting includes a multiple-interaction simulation. It 
was verified that exclusion of multiple interactions from the parton-level of {\sc Pythia}
results in a negligible change in the hadronisation corrections.

\section{Monte Carlo simulation
\label{sec:Monte-Carlo-simulations}}

The measured cross sections were compared to leading-order
Monte Carlo (MC) models
which use the QCD parton shower approach
to incorporate high-order QCD effects followed by fragmentation
into hadrons. The MC events were generated with
the {\sc Pythia} 6.3~\cite{cpc:46:43,cpc:82:74,cpc:135:238}
and with the {\sc Herwig}  6.5~\cite{cpc:67:465,jhep:01:010} models using the
default parameters in each case. The CTEQ5L \cite{epj:c12:375}
proton PDF was used together with the SaS-2D parameterisation \cite{pl:b376:193}
for the photon PDF.
Both direct and resolved prompt-photon events were generated.

The same MC samples were used to calculate the acceptance and
to evaluate the signal and background content of the sample.
Samples of background photoproduction events (without prompt-photon subprocesses) were
generated in addition to the prompt photon samples.
Both direct and resolved processes were simulated.
These MC samples  provided  background photons from the decay of hadrons
(predominantly from $\pi^0$ mesons).

The generated events were passed through a full simulation of the
detector using {\sc Geant} 3.13
\cite{tech:cern-dd-ee-84-1} and processed with the same reconstruction
program as used for the data.
The MC samples after the detector simulation do not give a good description of 
the $E_T$ and $\eta$ distributions
seen in the data. Such discrepancies are most prominent
at low $E_T^{\gamma}$, and were attributed to an inadequacy of the MC models.
For the acceptance calculations,
the MC distributions were reweighted to match the distributions in $E_T$ and $\eta$ of the data.
The reweighting  was
performed in four-dimensional phase space in $E_T$ and $\eta$ of the photon and of the
accompanying jet; thus correlations between these kinematic variables were properly
taken into account.

\section{Data selection and prompt-photon reconstruction}
\label{data}

\subsection{Event selection and jet reconstruction
\label{jets}}

The online selection made use of a standard ZEUS electron
finding algorithm to select events with an electromagnetic cluster~\cite{pl:b413:201}.
For the offline analysis, neutral-current
deep inelastic (DIS)  events with an
identified scattered electron candidate
were removed from the sample. This restricted the
virtuality of the incident photon  to the range $Q^2<1\gev^2$.
In addition, the following cuts were applied:

\begin{itemize}

\item
charged current DIS events were rejected by
requiring the total missing transverse momentum in the calorimeter
to be less than  $10\gev$;

\item
$0.2<y_{\rm JB}<0.8$, where
$y_{\rm JB}$ is
the inelasticity parameter reconstructed with the
Jacquet-Blondel method \cite{proc:epfacility:1979:391};

\item
$\mid Z_{\mathrm{vertex}} \mid  \le  50$ cm, where
$Z_{\mathrm{vertex}}$ is the event-vertex position
determined from the
tracks.

\end{itemize}

Jets were reconstructed by running the longitudinally invariant
$k_{T}$ algorithm  in inclusive mode \cite{pr:d48:3160,*np:406:187}
on energy-flow objects (EFOs) \cite{briskin:phd:1998}, which are based
on a combination of track and calorimeter information.
The jet variables $E_T$ and $\eta$ 
were defined according to the Snowmass convention \cite{proc:snowmass:1990:134}.  
Each jet was  classified as
either a photon candidate or a hadronic jet.
The photon-candidate jet was required to consist of  EFOs without  associated tracks
and to be within the CTD and BCAL acceptance,
 $-0.74<\eta^{\gamma}<1.1$.
For this jet,
$E_{\rm EMC}/E_{\rm tot}>0.9$ is  required, where  $E_{\rm EMC}$ is
the energy reconstructed in the electromagnetic part of the CAL and
$E_{\rm tot}$ is the total energy of this jet. 
After correction for energy losses, the cut $E_{T}^{\gamma}>5\gev$
was applied.

Hadronic jets, after correction for energy losses,
were  selected in the kinematic range $E_{T}^{\rm jet}>6\gev$, 
$-1.6<\eta^{\rm jet}<2.4$. They were required to have  $E_{\rm EMC}/E_{\rm tot}<0.9$.
If more than one jet was found within the above kinematic cuts,
the jet with the highest $E_{T}^{\rm jet}$ was accepted.
The minimum transverse momentum of
the hadronic jet was set to be higher than for the photon candidate, since
the NLO calculations employed in this analysis
are unstable for symmetric cuts on the minimum
transverse momenta of both jets \cite{np:b507:315,Fontannaz:2001nq}.

In total, 3910 events with a prompt-photon candidate and a 
hadronic jet were  selected.

\subsection{Identification of isolated photons and hadrons
\label{BPREmeth}}

For the prompt-photon identification,
the conversion-probability method based on the BPRE was used.
In contrast to the shower-profile approach used in previous measurements 
\cite{pl:b413:201,pl:b511:19,pl:b472:175},  the present
approach uses the probability
of conversion of photons to  $e^{+}e^{-}$ pairs in 
detector elements and inactive material,  mainly the solenoid located
in front of the BCAL.
The conversion probability for a single photon is
lower than for multiphoton events arising from neutral meson decays ($\pi^0$, $\eta$, etc.); therefore, small BPRE signals can be used to identify isolated photons.

The response of the BPRE to single isolated photons was studied
using the deeply virtual 
Compton scattering (DVCS) data, $ep \to e^{'} \gamma p$,  taken during 1999-2000.
This sample is known to provide photons of high purity \cite{ZEUScompton}.
Events with two isolated electromagnetic clusters and one CTD
track were pre-selected.
One cluster was required to have energy $E_{e'}>8\gev$ and to
be associated with the CTD track,
thereby ensuring compatibility with the scattered electron.
The cluster without an associated track was then reconstructed
using the
$k_T$ cluster algorithm as described in Section~\ref{jets}.
The photon candidate was  required to be
in the BCAL region,  $-0.74<\eta^{\gamma}<1.1$,
and to have energy in the range $5<E_{\gamma}<10\gev$.
The BPRE signal for the DVCS photons
was determined as the sum of the signal of the BPRE tiles
whose centre falls within a cone of size $0.7$ in  $\eta$-$\phi$ around the
photon candidate.
A smaller cone size leads to an efficiency that is not well
reproduced by the MC.
Other details of the DVCS selection and MC  simulations are
given elsewhere \cite{ZEUScompton,misc:gendvcs}.

In the DVCS sample, the fraction of events with BPRE signal below one mip
is very sensitive to the amount of material in front of the BPRE
since such events are dominated by non-converted photons.
The MC simulation overestimates this fraction by
$19\%$ compared to the data due to
an inadequate simulation of the material in front of the BPRE.
This discrepancy does not have a significant dependence
on the cone size.

The amount of inactive material was further studied using scattered electrons from DIS
events.  This study indicated that more material
in front of the BCAL was necessary for the MC simulation.

Using a dedicated {\sc Geant}
simulation~\cite{tech:cern-dd-ee-84-1}, it was found that an
increase of inactive material in the MC simulation by $0.25{\rm X_{0}}$ 
was sufficient to describe the fraction of events without
photon conversions seen in the DVCS data.
For this {\sc Geant} simulation, it
was assumed that all inactive material is distributed
uniformly in the region of the ZEUS solenoid
located in front of the BPRE tiles.
The effect of the additional material was then taken into
account in the standard ZEUS MC by applying a correction to the
BPRE distribution based on the results of the dedicated simulation.

Figure~\ref{prph_1} shows the BPRE signal for the
DVCS data compared to the DVCS Monte Carlo model~\cite{misc:gendvcs}
after correction for additional material. 
There is good agreement between the  data and the MC distribution.
This shows that the inactive  material and the BPRE  resolution
are well represented in the MC simulation.

The MC response of the BPRE to single hadrons was studied for $\pi^0$ and $\eta$ mesons.
Since these mesons decay
to several photons, more conversions to $e^+e^-$ will occur
than for single photons.
As expected, the average BPRE
signal for $\pi^0$ and $\eta$ mesons
was larger  than for the isolated photons.
An example of the MC simulation of the BPRE response to
single $\gamma$, $\pi^0$ and $\eta$ is shown in Figure~\ref{prph_1a}.
The BPRE  distributions for $\pi^0$ and $\eta$ mesons
were also corrected to take into account
additional dead material in front of the BPRE.

\subsection{Extraction of the prompt-photon signal
\label{BPREmeth2}}

The BPRE signal for prompt-photon candidates selected as
described in Sect.~\ref{jets} was determined using a cone of radius $0.7$
in $\eta$-$\phi$ space, as was done for the DVCS analysis.
Figure~\ref{prph_2} shows the comparison  between the data and the {\sc Pythia}  MC for:
the BPRE signal for the photon candidate;
the difference between the total calorimeter energy and the energies of the jet and
the photon candidate, $\Delta E=E_{\rm tot}-E^{\rm jet}-E^{\gamma}$;
and the distance from the photon candidate to any EFO in an event,
$$
D=\sqrt{{(\eta_{\gamma}-\eta_{\mathrm{EFO}})^{2}+(\phi_{\gamma}-\phi_{\mathrm{EFO}})^{2}}},
$$
where $\eta_{\gamma}$ ($\eta_{\mathrm{EFO}}$)  and $\phi_{\gamma}$
($\phi_{\mathrm{EFO}}$)
are the azimuthal angle and pseudorapidity of
the photon candidate (EFO).

Figure~\ref{prph_2}(a) shows that there is a significant fraction of events
with a small number of mips, similar to the DVCS data.
However,  since
the dijet photoproduction cross section is
higher by several orders of magnitude than the $\gamma$+jet cross section,
there is additional hadronic background
even after the cuts discussed in Section~\ref{jets}.

The BPRE distribution for the prompt-photon candidates
was used to determine the background fraction.
The fraction of inclusive dijet photoproduction events needed was found
from a $\chi^2$-minimisation  procedure.
After the inclusion of the background events,
the shape of the BPRE distribution for the prompt-photon candidates
is well reproduced by the MC simulation, as shown in Figure~\ref{prph_2}(a).

The inclusion of the dijet background  leads to good  description of the CAL
distributions shown in  Fig.~\ref{prph_2}(b) and (c), which are also sensitive
to the prompt-photon events.
On average, the $\Delta E$  should be larger for the dijet events, where
more energy is radiated outside the dijet system than for the $\gamma$+jet events.
After the inclusion of the background,
this distribution is well reproduced by MC.

The background fraction described above
was used in the calculation of the total $\gamma$+jet  cross section.
For differential cross sections, the background fractions were determined
by fitting the BPRE signal independently in each bin of the respective distributions.
In order to reduce statistical fluctuations in regions of small statistics,
it was assumed that the background fractions varied smoothly from bin to bin.
Therefore, the dependence of the background
fraction on  $E_{T}^{\gamma}$, $\eta^{\gamma}$,
$E_{T}^{\rm jet}$, $\eta^{\rm jet}$ and $x_{\gamma}^{\rm obs}$  was obtained by fitting
the background fractions for  each bin with a linear function.
The number of prompt-photon events in each kinematic bin was determined from
such a linear-regression fit. The statistical uncertainties on the
number of signal events were evaluated using $68\%$ confidence-level limits on the linear
fit of the fractions.

\section{Cross section calculations  and systematic \\ uncertainties}
\label{syst}

The differential cross sections for a given observable $Y$ were
determined  as:
$$
\frac {d\sigma}{dY} = \frac {N } {A \cdot \mathcal {L} \cdot
\Delta Y} \>\> ,
$$
where $N$ is the number of prompt-photon events
in a bin of size $\Delta Y$,
$A$ is the acceptance
and $\mathcal {L}$ is the integrated luminosity.
The acceptance was calculated
using MC from the ratio of 
the number of reconstructed events after the selection cuts  to the
number of generated events.

The systematic uncertainties
were evaluated by
changing the selection and the analysis procedure.  The
contribution of each cut variation  to the total cross section is
given in parentheses as a percentage of the total cross section:

\begin{itemize}

\item
the calorimeter energy scale was changed by $\pm 3\%$ ($^{+9.1}_{-11.7}\%$);

\item
the transverse momentum cut and the $\eta$ range for the photon and hadron jet
were lowered (raised) independently by one $\sigma$ of the
resolution. The systematic  uncertainty due to the transverse-energy cut
for the photon was found to be ($^{+2.5}_{-3.7}\%$).
The largest systematical uncertainty due to the transverse-energy cut
on the jet was ($^{+2.2}_{-2.0}\%$).
The systematic uncertainty associated with
the variations in the pseudorapidity was small ($\pm 0.8\%$);

\item
the uncertainty in the quantity of inactive material in the MC
was estimated by varying the amount of inactive material by $\pm 5\%$ of a ${\rm X_0}$
in the {\sc Geant}-based correction factors ($^{-7.0}_{+5.0}\%$);

\item
the cone radius for the determination of the BPRE signal was
changed by $\pm 0.1$ units ($^{-2.3}_{+2.7}\%$).
A larger cone size leads to a larger leakage of hadronic energy into the photon, 
which is not well simulated in MC;

\item
variations of the cuts on $y_{\rm JB}$,  $Z_{\mathrm{vertex}}$
and  on total missing transverse energy ($\pm 2\%$);

\item
the resolved contribution in MC
was changed by $\pm 15\%$ ($<1\%$);

\item
the cut on the electromagnetic fraction $E_{\rm EMC}/E_{\rm tot}$ for the photon jet
was changed by $\pm 0.02$ ($^{+0.9}_{-1.5}\%$);

\item
the acceptance correction and the fraction of background photoproduction
events was determined using {\sc Herwig} ($-0.6\%$).

\end{itemize}

The overall systematic uncertainty was determined by adding the above
uncertainties in quadrature. A $2\%$ normalisation uncertainty  due to
the luminosity measurement error was not included in the systematic
uncertainties.

As an additional check, the differential $\gamma$+jet cross sections were
found by using the global background fraction determined in the full kinematic range.
Further, the cross sections were calculated
from the number of the detector-level events in the data and MC after requiring a
BPRE signal $<7$ mip, i.e. in a region where the purity of the prompt photon sample
is expected to be above  $50\%$.
The results from these alternative methods were 
consistent with the final cross sections.

\section{Results}
\label{result}

The total cross section for the process $ep\to e+\gamma_{\rm prompt}+\mathrm{jet}+X$
for $0.2<y<0.8$, $Q^{2}<1\,\mathrm{GeV^{2}}$, $5<E_{T}^{\gamma}<16$
GeV, $6<E_{T}^{\rm jet}<17$ GeV, $-0.74<\eta^{\gamma}<1.1$,  $-1.6<\eta^{\rm jet}<2.4$
and  $E_{T}^{\gamma, \rm (true)}>0.9\, E_{T}^{\gamma}$  was measured to be  

$$
\sigma(ep\to e+\gamma_{\rm prompt}+\mathrm{\rm jet}+X)=33.1\pm 3.0\,(\mathrm{stat.})\,_{-4.2}^{+4.6}(\mathrm{syst.})
\:\mathrm{pb.}
$$

This cross section should be compared to the QCD predictions 
after the hadronisation corrections:
$23.3^{+1.9}_{-1.7}\pb$ (KZ),
$23.5^{+1.7}_{-1.6}\pb$ (FGH)  and $30.7^{+3.2}_{-2.7}\pb$ (LZ).
The scale uncertainties on the QCD calculations were estimated
by varying $\mu_R$  between $\mu_R /2$  and $ 2 \mu_R$.
The {\sc Pythia} and {\sc Herwig} cross sections are $20.0\pb$ and  $13.5\pb$,
respectively.

The differential cross sections as functions of $E_{T}$ and $\eta$ for the prompt-photon
candidates and for the accompanying jets are shown in Figs.~\ref{prph_3}
and \ref{prph_4}.
Figure~\ref{prph_5}  shows the  distribution for
$x_{\gamma}^{\rm obs}$  defined as $\sum_{\gamma, \rm jet} (E_i-P_Z^i)/(2E_e y)$
(the sum runs over the photon candidate and the hadronic jet).
Table~\ref{table} gives the differential cross sections with 
the statistical and systematical
uncertainties,
as well as the hadronisation-correction factors calculated
in the same bins as the data\footnote{The actual
hadronisation corrections applied to the NLO calculations
shown in  Figs.~\ref{prph_3}-\ref{prph_7} were calculated using finer bins.}.

The {\sc Pythia} and {\sc Herwig}  differential cross sections do not rise as steeply
at low $E_{T}^{\gamma}$ as do the data. In addition, they
underestimate the measured cross sections.
The  KZ NLO prediction,  corrected for hadronisation
effects as described in Sect.~\ref{sec:NLO},
describes the data better.
However,  it underestimates the observed cross section at low
$E_{T}^{\gamma}$ and in the forward jet region.
The observed difference between the data and
the NLO QCD calculations
is concentrated  in the $x_{\gamma}^{\rm obs} < 0.75$ region which
is sensitive to the resolved photon contribution.

The FGH prediction is similar to the KZ NLO.
The largest difference between the two predictions is found for the
$\eta^{\rm jet}$ cross section, where the FGH cross section is  closer to the data
in the forward jet region.
The renormalisation scale uncertainty for
the FGH QCD calculations is similar to that estimated for the KZ predictions
(not shown).

The LZ prediction based on the $k_T$-factorisation approach corrected
for hadronisation effects  gives the best description
of the $E_{T}$ and $\eta$ cross sections.
In particular, it describes the lowest $E_{T}^{\gamma}$ region better than  the
KZ and FGH NLO predictions.
The $\eta^{\rm jet}$ cross section for the
associated jet in the forward region is also better reproduced by the LZ calculation.

It is difficult to compare the present cross sections with the H1
result~\cite{H1prompt}, since a significant model-dependent extrapolation
to the low $E_{T}^{\rm jet}$ region used by H1 is required.
A comparison in the region $E_{T}^{\rm jet}>6\gev$ shows good agreement
with H1 for the $E_{T}^{\rm jet}$ differential cross section.

Since the largest difference between the NLO calculations and the data is concentrated
in the region of low  $E_{T}^{\gamma}$ and low  $E_{T}^{\rm jet}$, it is instructive
to verify the level of agreement with NLO when the minimum transverse energy of the detected
prompt photons is increased from $5\gev$ to $7\gev$.
In this case, hadronisation corrections are expected to be smaller.
Further, in comparison with the previous measurements,
such a choice may emphasize different aspect of contributions 
of high-order QCD radiation~\cite{Fontannaz:2001nq}, 
since the transverse energy of the prompt-photon is larger
than that of the jet.

The total $ep\to e+\gamma_{\rm prompt}+\mathrm{jet}+X$  cross section
for $E_{T}^{\gamma}>7\gev$ (keeping the other cuts the same as before)
is
$\sigma=13.8\pm 1.2\,(\mathrm{stat.})\,_{-1.6}^{+1.8}(\mathrm{syst.})
\:\mathrm{pb}$.
This result agrees with the QCD calculations after the hadronisation corrections:
$14.9^{+1.3}_{-1.0}\pb$ (KZ),
$13.4^{+1.1}_{-0.9}\pb$ (FGH)  and $13.6^{+0.9}_{-1.0}\pb$ (LZ).
The PYTHIA and HERWIG models predict $13.7\pb$ and $9.4\pb$, respectively.

Figures~\ref{prph_6}, \ref{prph_7}  and Table~\ref{table_new} show the corresponding
differential cross
sections. The applied hadronisation corrections are given in Table~\ref{table_new}.
For the $E_{T}^{\gamma}>7\gev$ cut, both the NLO QCD and the LZ predictions
agree well with the data.
The {\sc Pythia} MC model also agrees  well with the
cross sections, while {\sc Herwig} is still below the data.

\section{Conclusions}

The photoproduction of prompt photons, together with an
accompanying jet, has been measured in $ep$ collisions at a centre-of-mass energy of
318 GeV with the ZEUS detector at HERA using
an integrated luminosity of 77  pb$^{-1}$.

In the kinematic region $E_{T}^{\gamma}>5\gev$ and $E_{T}^{\rm jet}>6\gev$
the prompt-photon data disagree with
the available MC predictions which predict a less steep rise of the cross sections with 
decreasing $E_{T}^{\gamma}$.
The discrepancy is reduced for the NLO calculations.
However, they still underestimate the data in the low 
$E_T^{\gamma}$ and $E_T^{\rm jet}$ regions, which 
are likely to be the most sensitive to
the treatment of high-order QCD terms and hadronisation effects.
The best description of the data
was found for the calculations based on the $k_T$-factorisation approach and
unintegrated parton densities.

When the minimum transverse energy of prompt photons is increased 
from $5\gev$ to
$7\gev$, both NLO QCD and the $k_T$-factorisation calculations describe the data well.

\section*{Acknowledgements}
\vspace{0.3cm}
We thank the DESY Directorate for their strong support and encouragement.
The remarkable achievements of the HERA machine group were essential for
the successful completion of this work and are greatly appreciated. We
are grateful for the support of the DESY computing and network services.
The design, construction and installation of the ZEUS detector have been
made possible owing to the ingenuity and effort of many people from DESY
and home institutes who are not listed as authors.
We thank M.~Fontannaz, G.~Heinrich, M.~Krawczyk, A.~Lipatov
and A.~Zembrzuski for discussions and
for providing the QCD calculations.

\vfill\eject

%% file: DESY-06-125-ref.tex
{
\def\bibname{\Large\bf References}
\def\refname{\Large\bf References}
\pagestyle{plain}
\ifzeusbst
  \bibliographystyle{./BiBTeX/bst/l4z_default}
\fi
\ifzdrftbst
  \bibliographystyle{./BiBTeX/bst/l4z_draft}
\fi
\ifzbstepj
  \bibliographystyle{./BiBTeX/bst/l4z_epj}
\fi
\ifzbstnp
  \bibliographystyle{./BiBTeX/bst/l4z_np}
\fi
\ifzbstpl
  \bibliographystyle{./BiBTeX/bst/l4z_pl}
\fi
{\raggedright
\bibliography{./BiBTeX/user/syn.bib,%
              ./BiBTeX/bib/l4z_articles.bib,%
              ./BiBTeX/bib/l4z_books.bib,%
              ./BiBTeX/bib/l4z_conferences.bib,%
              ./BiBTeX/bib/l4z_h1.bib,%
              ./BiBTeX/bib/l4z_misc.bib,%
              ./BiBTeX/bib/l4z_old.bib,%
              ./BiBTeX/bib/l4z_preprints.bib,%
              ./BiBTeX/bib/l4z_replaced.bib,%
              ./BiBTeX/bib/l4z_temporary.bib,%
              ./BiBTeX/bib/l4z_zeus.bib}}
}
\vfill\eject

%% file: DESY-06-125-tab.tex

\vspace{2.0cm}

\begin{table}
\centering
   \input{DESY-06-125_table_1}
\caption{
The differential prompt-photon cross sections with additional jet 
requirement measured in the region  
$0.2<y<0.8$, $Q^{2}<1\gev^2$, $5<E_{T}^{\gamma}<16\gev$, 
$6<E_{T}^{jet}<17\gev$, $-0.74<\eta^{\gamma}<1.1$ and $-1.6<\eta^{jet}<2.4$.
The statistical and systematical uncertainties are given separately.
The hadronisation correction factors (see the text) applied to the QCD
calculations for the same kinematic bins as for the data are also shown.  
}
\label{table}

\end{table}

\begin{table}
\centering
   \input{DESY-06-125_table_2}
\caption{
The differential prompt-photon cross sections with additional jet
requirement measured in the region defined
as for Table~\ref{table}
except for the cut on the
transverse energy of the prompt photon, which was increased to $7\gev$.
The statistical and systematical uncertainties are given separately.
The hadronisation correction factors applied to the QCD 
calculations for the same kinematic bins as for the data are also shown.
}
\label{table_new}

\end{table}

\vfill\eject

%% file: DESY-06-125_table_1.tex
 \begin{tabular}{|c|c|c|}
 \hline
 $E_{T}^{\gamma}$ (GeV) & $d\sigma /dE_{T}^{\gamma}$ (pb/GeV) & $C^{\rm had}$ \\
 \hline
 $ 5.00, 7.00$ & $  9.0\pm   1.1^{ +  1.6}_{ -  1.7}$ & 0.78 \\ \hline
 $ 7.00, 9.00$ & $  3.7\pm   0.6^{ +  0.9}_{ -  0.6}$ & 1.01 \\ \hline
 $ 9.00, 11.00$ & $  2.9\pm   0.5^{ +  0.3}_{ -  0.4}$ & 1.05 \\ \hline
 $11.00, 13.00$ & $  0.7\pm   0.2^{ +  0.2}_{ -  0.2}$ & 1.06 \\ \hline
 $13.00, 16.00$ & $  0.3\pm   0.1^{ +  0.1}_{ -  0.1}$ & 1.06 \\ \hline
 \hline
 $\eta^{\gamma}$ & $d\sigma /d\eta^{\gamma}$ (pb)  & $C^{\rm had}$ \\
 \hline
 $-0.74, -0.34$ & $ 21.7\pm   3.3^{ +  3.6}_{ -  3.7}$  &  0.89 \\ \hline
 $-0.34,  0.02$ & $ 24.0\pm   3.3^{ +  3.5}_{ -  2.9}$ &  0.91 \\ \hline
 $ 0.02, 0.38$ & $ 21.5\pm   3.0^{ +  3.4}_{ -  3.2}$  &  0.93 \\ \hline
 $ 0.38, 0.74$ & $ 16.3\pm   2.3^{ +  2.2}_{ -  3.5}$ &  0.95 \\ \hline
 $ 0.74, 1.10$ & $ 12.7\pm   6.4^{ +  4.3}_{ -  4.2}$ &  0.95 \\ \hline
 \hline
 $E_{T}^{\rm jet}$ (GeV) & $d\sigma /dE_{T}^{\rm jet}$ (pb/GeV)  & $C^{\rm had}$ \\
 \hline
 $ 6.00, 8.00$ & $ 10.9\pm   1.2^{ +  1.5}_{ -  1.8}$ &  0.89 \\ \hline
 $ 8.00, 10.00$ & $  3.1\pm   0.5^{ +  0.6}_{ -  0.4}$ &  0.96 \\ \hline
 $10.00, 12.00$ & $  2.0\pm   0.4^{ +  0.3}_{ -  0.3}$ &  0.97 \\ \hline
 $12.00, 14.00$ & $  1.3\pm   0.4^{ +  0.2}_{ -  0.2}$ &  0.95 \\ \hline
 $14.00, 17.00$ & $  0.6\pm   0.2^{ +  0.1}_{ -  0.0}$ &  0.90 \\ \hline
 \hline
 $\eta^{\rm jet}$ & $d\sigma /d\eta^{\rm jet}$ (pb)  & $C^{\rm had}$ \\ \hline
 $-1.60, -0.80$ & $  3.3\pm   0.9^{ +  0.9}_{ -  0.7}$  &  0.74 \\ \hline
 $-0.80,  0.00$ & $ 11.7\pm   1.4^{ +  1.7}_{ -  1.3}$ &  0.85 \\ \hline
 $ 0.00,  0.80$ & $  8.9\pm   1.5^{ +  1.4}_{ -  1.5}$ &  0.99 \\ \hline
 $ 0.80,  1.60$ & $  8.3\pm   2.2^{ +  2.3}_{ -  2.4}$ &  1.07 \\ \hline
 $ 1.60,  2.40$ & $ 10.7\pm   2.0^{ +  1.4}_{ -  2.4}$ &  1.09 \\ \hline
 \hline
 $x^{\mathrm{obs}}_{\gamma}$  & $d\sigma /dx^{\mathrm{obs}}_{\gamma}$ (pb)  & $C^{\rm had}$ \\ \hline
 $ 0.00,  0.25$ & $  3.9\pm   3.1^{ +  4.2}_{ -  4.2}$ & 0.91 \\ \hline
 $ 0.25,  0.50$ & $ 37.9\pm   7.9^{ +  8.8}_{ - 12.3}$ & 0.95 \\ \hline
 $ 0.50,  0.75$ & $ 24.5\pm   5.2^{ +  6.3}_{ -  9.2}$  & 1.06 \\ \hline
 $ 0.75,  1.00$ & $ 80.4\pm   7.2^{ +  9.4}_{ - 12.6}$ & 0.90 \\ \hline
 \end{tabular}

%% file: DESY-06-125_table_2.tex
 \begin{tabular}{|c|c|c|}
 \hline
 $\eta^{\gamma}$ & $d\sigma /d\eta^{\gamma}$ (pb) &  $C^{\rm had}$ \\
 \hline
 $-0.74, -0.34$ & $  5.0\pm   0.9^{ +  1.1}_{ -  0.7}$ &  0.99 \\ \hline
 $-0.34,  0.02$ & $  8.2\pm   1.3^{ +  1.6}_{ -  1.4}$ &  1.00 \\ \hline
 $ 0.02,  0.38$ & $  9.0\pm   1.4^{ +  1.4}_{ -  1.2}$ &  1.02 \\ \hline
 $ 0.38, 0.74$ & $  7.9\pm   1.6^{ +  1.2}_{ -  1.3}$ &  1.04 \\ \hline
 $ 0.74, 1.10$ & $  8.0\pm   2.7^{ +  0.7}_{ -  2.7}$ &  1.05 \\ \hline
 \hline
 $E_{T}^{\rm jet}$ (GeV) & $d\sigma /dE_{T}^{\rm jet}$ (pb/GeV) &  $C^{\rm had}$ \\
 \hline
 $ 6.00,  8.00$ & $  2.5\pm   0.4^{ +  0.4}_{ -  0.4}$ &  1.08 \\ \hline
 $ 8.00, 10.00$ & $  2.0\pm   0.4^{ +  0.3}_{ -  0.2}$ &  1.00 \\ \hline
 $10.00, 12.00$ & $  1.3\pm   0.2^{ +  0.2}_{ -  0.2}$ &  0.99 \\ \hline
 $12.00, 14.00$ & $  0.5\pm   0.2^{ +  0.1}_{ -  0.1}$ &  0.97 \\ \hline
 $14.00, 17.00$ & $  0.2\pm   0.1^{ +  0.1}_{ -  0.0}$ &  0.91 \\ \hline
 \hline
 $\eta^{\rm jet}$  & $d\sigma /d\eta^{\rm jet}$ (pb)  &  $C^{\rm had}$ \\ \hline
 $-1.60, -0.80$ & $  1.1\pm   0.3^{ +  0.3}_{ -  0.4}$ &  0.85 \\ \hline
 $-0.80,  0.00$ & $  4.2\pm   0.7^{ +  0.4}_{ -  0.5}$ &  0.95 \\ \hline
 $ 0.00,  0.80$ & $  5.6\pm   1.0^{ +  0.8}_{ -  0.6}$ &  1.06 \\ \hline
 $ 0.80,  1.60$ & $  3.4\pm   0.8^{ +  0.8}_{ -  0.6}$ &  1.15 \\ \hline
 $ 1.60,  2.40$ & $  1.8\pm   0.5^{ +  0.6}_{ -  0.5}$ &  1.18 \\ \hline
 \hline
 $x^{\mathrm{obs}}_{\gamma}$  & $d\sigma /dx^{\mathrm{obs}}_{\gamma}$ (pb) &  $C^{\rm had}$ \\ \hline
 $ 0.00,  0.25$ & $  1.6\pm   0.9^{ +  2.1}_{ -  1.7}$ & 1.15 \\ \hline
 $ 0.25,  0.50$ & $  7.3\pm   1.5^{ +  2.0}_{ -  3.1}$ & 1.11 \\ \hline
 $ 0.50,  0.75$ & $  9.1\pm   1.6^{ +  2.5}_{ -  1.7}$ & 1.16 \\ \hline
 $ 0.75,  1.00$ & $ 33.9\pm   4.4^{ +  5.3}_{ -  5.5}$ & 0.99 \\ \hline
 \end{tabular}

%% file: DESY-06-125-fig.tex
\begin{figure}
\begin{center}

\begin{minipage}[c]{0.48\textwidth}
\includegraphics[height=4.5cm]{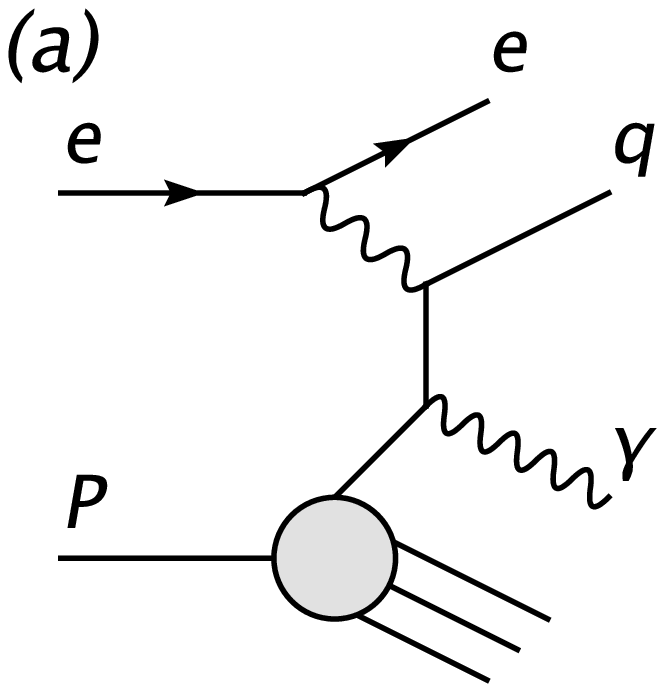}
\end{minipage}
\hfill
\begin{minipage}[c]{.48\textwidth}
\includegraphics[height=4.5cm]{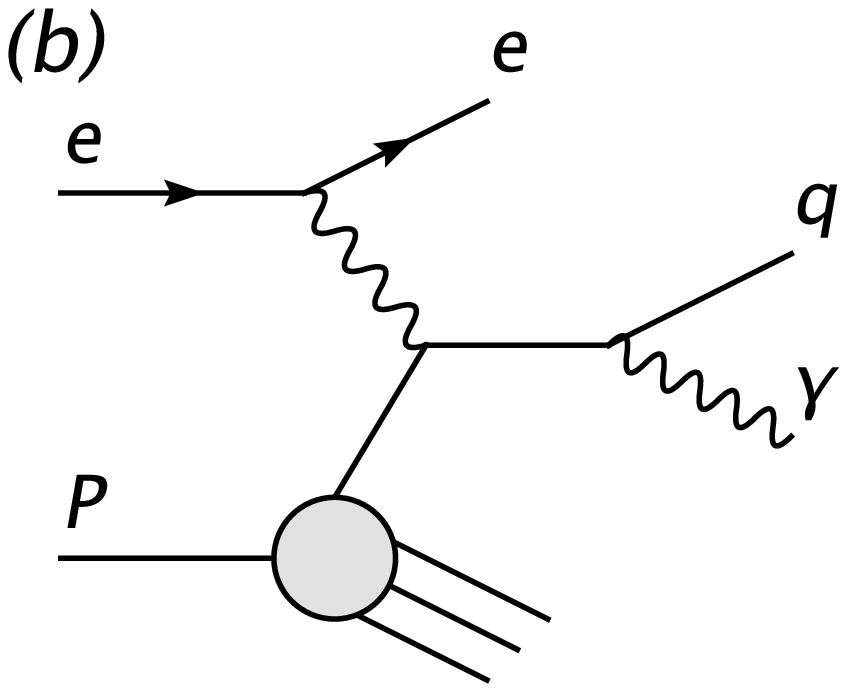}
\end{minipage}

\vspace{0.9cm}

\begin{minipage}[c]{0.48\textwidth}
\includegraphics[height=4.5cm]{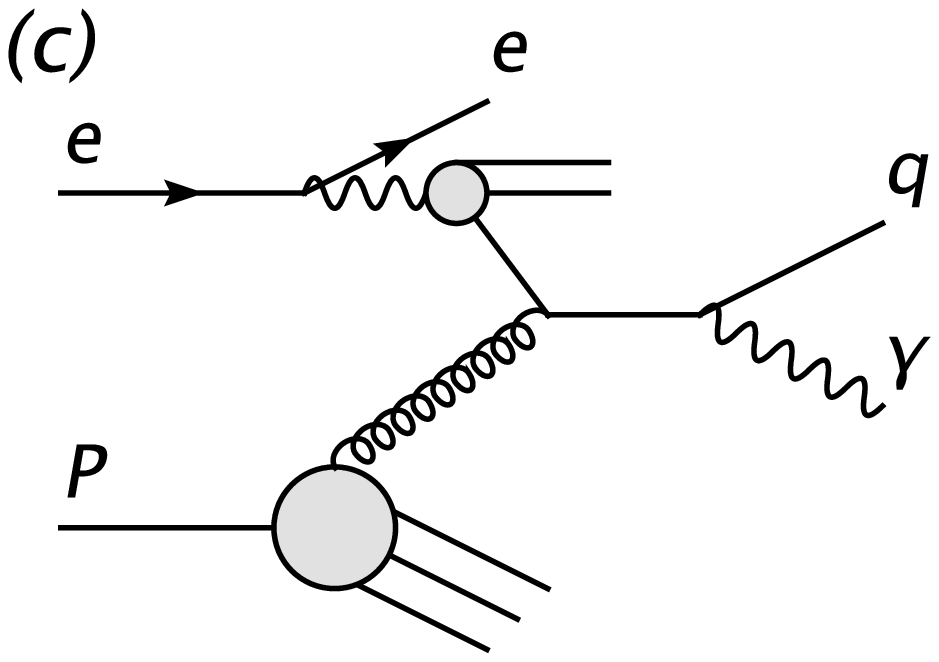}
\end{minipage}
\hfill
\begin{minipage}[c]{.48\textwidth}
\includegraphics[height=4.5cm]{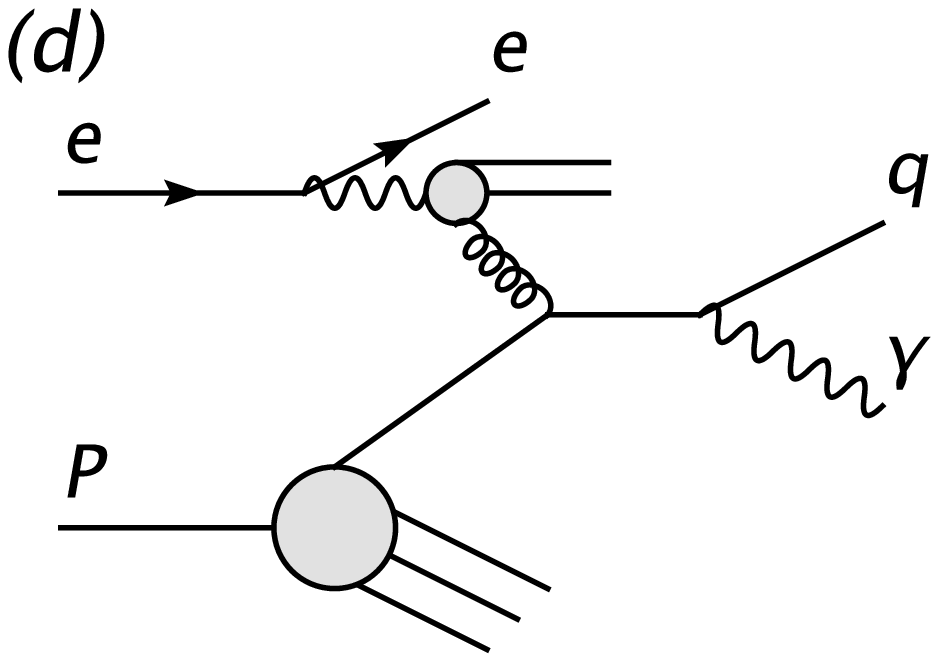}
\end{minipage}
\caption{Examples of diagrams for $\gamma$+jet events at leading order:  direct photon interactions, (a-b) and resolved photon interactions, (c-d). The resolved diagrams with t-channel exchange are not shown.
}
\label{prph_fe}
\end{center}
\end{figure}

\begin{figure}
\begin{center}
\includegraphics[height=16.0cm]{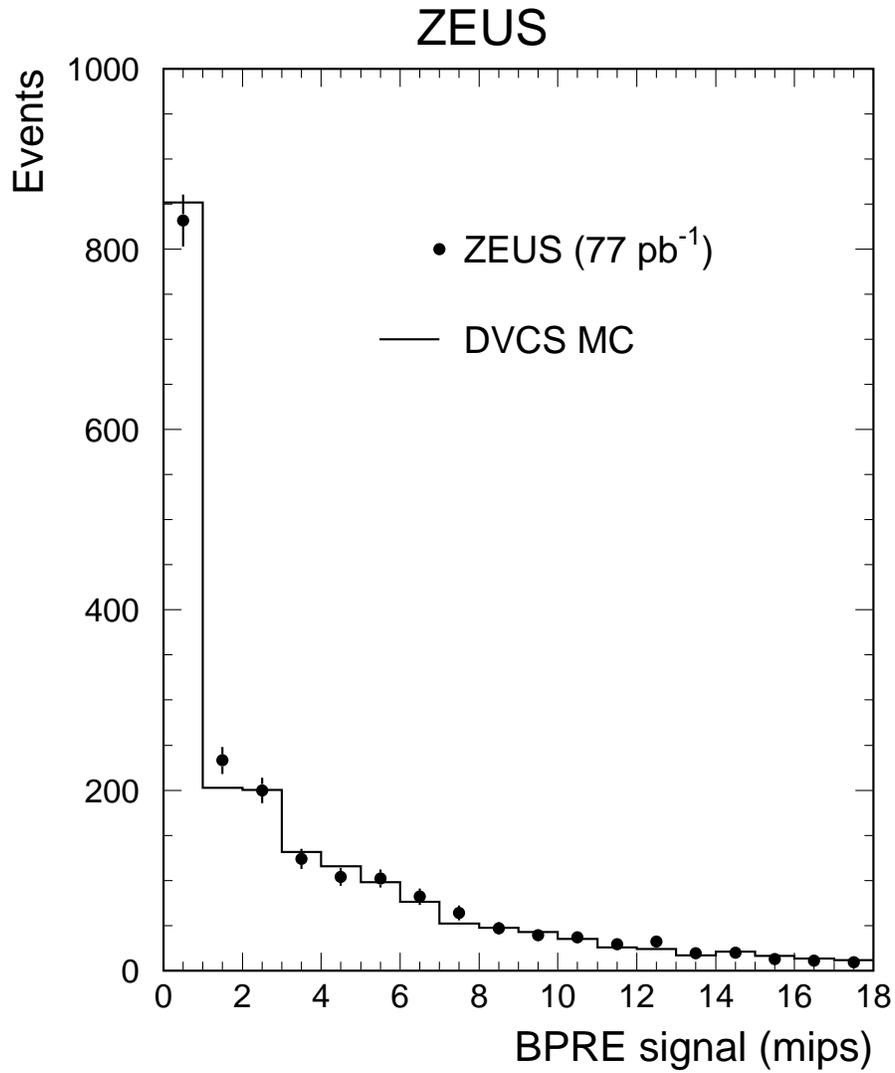}%
\caption{The response of the BPRE detector to isolated photons in the DVCS
data sample. The DVCS Monte Carlo distribution 
was normalised to the data.
}
\label{prph_1}
\end{center}
\end{figure}
 

\begin{figure}
\begin{center}
\includegraphics[height=10.0cm]{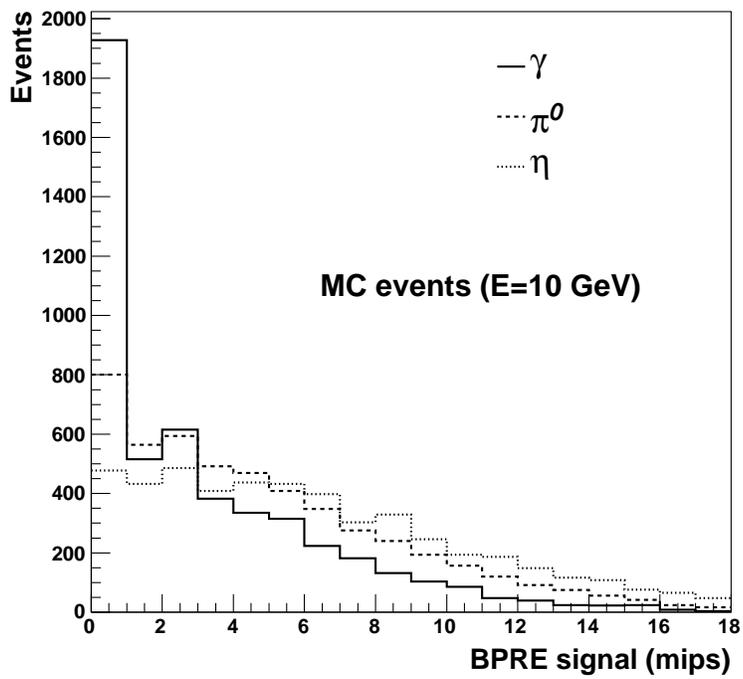}%
\caption{The BPRE response to isolated photons,  $\pi^0$ and $\eta$ in
the MC simulation.  
An initial energy of $10\gev$ for all particles was used. The amount of inactive
material in front of the BPRE was set to $1.25 X_0$.      
}
\label{prph_1a}
\end{center}
\end{figure}


\begin{figure}
\begin{center}
\includegraphics[height=18.0cm]{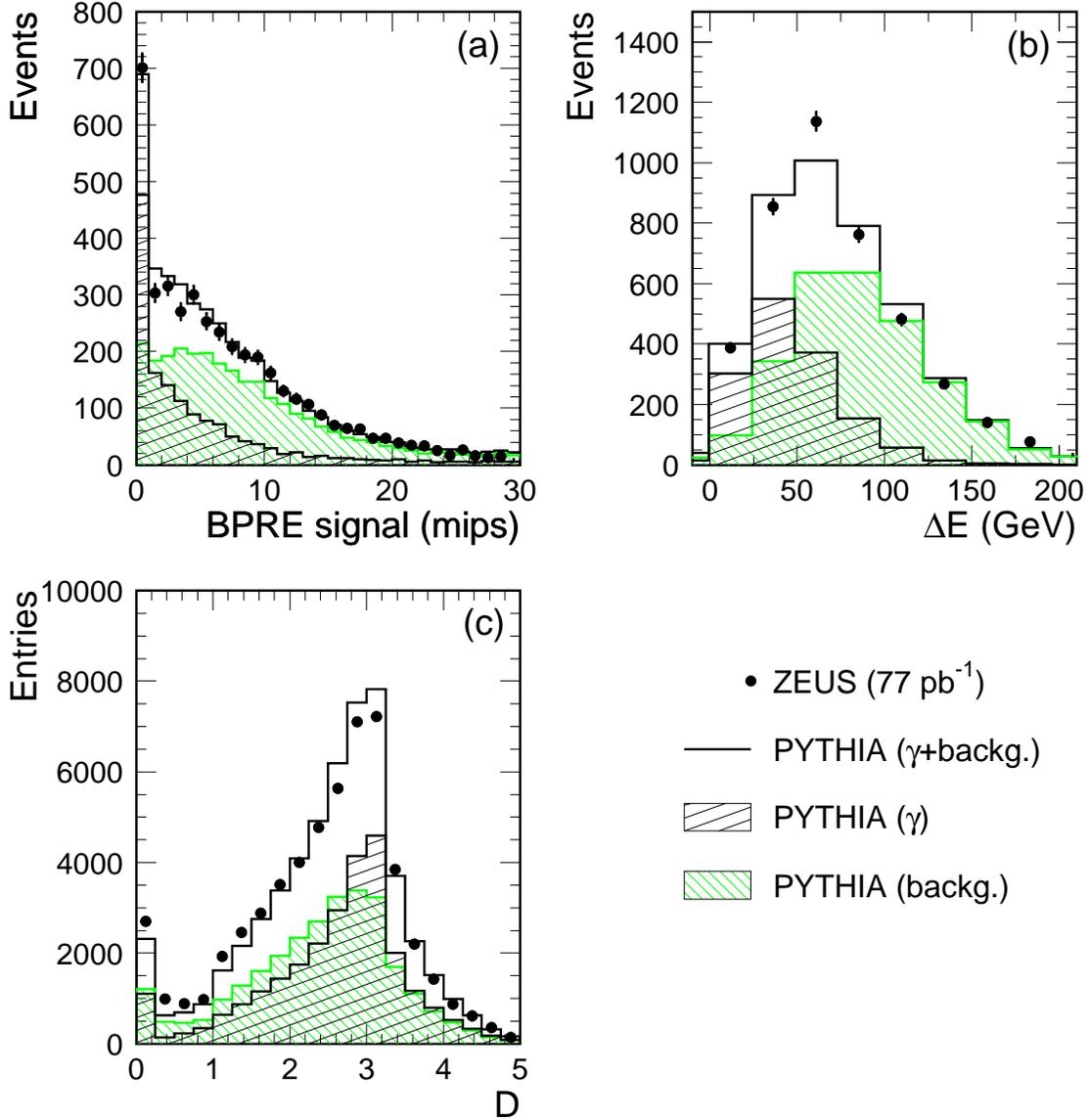}%
\caption{Comparisons between the data and MC for:
(a) the BPRE signal for the photon candidate;
(b) the difference between the total calorimeter energy and the energies of the jet and
the photon candidate, $\Delta E=E_{\rm tot}-E^{\rm jet}-E^{\gamma}$; and 
(c) the distance from the photon candidate to any EFO (see the text).
The non-hatched histogram is the sum of the prompt photon MC and the background MC.
The fraction of the prompt-photon events was found after a  $\chi^2$ minimisation
procedure for the BPRE distribution shown in (a). 
}
\label{prph_2}
\end{center}
\end{figure}
 

\begin{figure}
\begin{center}
\includegraphics[height=19.0cm]{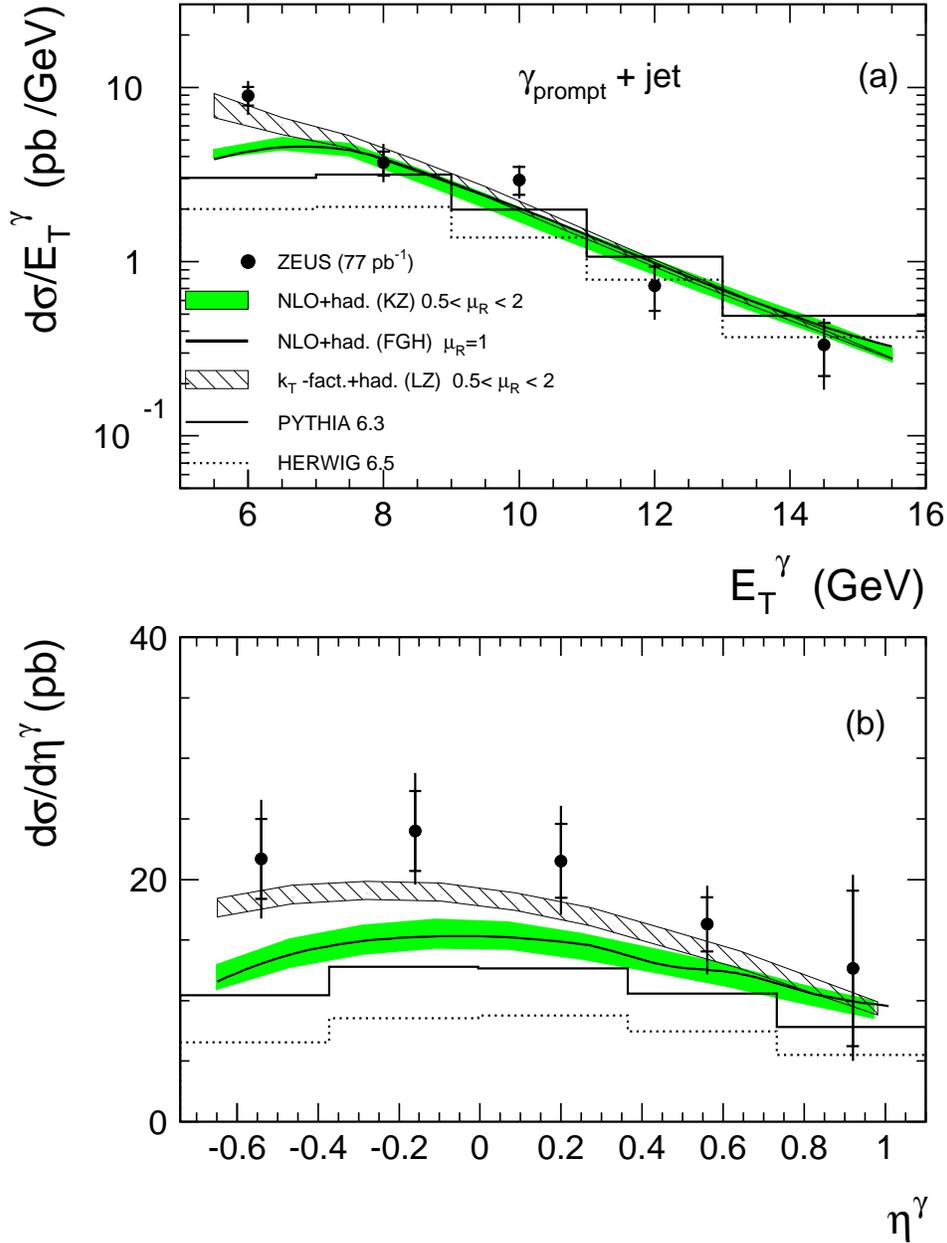}%
\caption{The  $\gamma+\mathrm{jet}$ differential cross sections
as functions of $E_{T}^{\gamma}$ and $\eta^{\gamma}$ 
compared to theoretical QCD calculations (with hadronisation corrections included). 
The histograms show the predictions of the Monte Carlo models.
The inner error bars show the statistical uncertainties,
the outer ones show statistical and systematic uncertainties added
in quadrature. The shaded bands for the KZ prediction correspond to the 
uncertainty in the 
renormalisation
scale which was changed by a factor of 0.5 and 2.
A similar uncertainty exists for the FGH prediction (not shown).  
} 
\label{prph_3}
\end{center}
\end{figure}

\newpage
\begin{figure}
\begin{center}
\includegraphics[height=19.0cm]{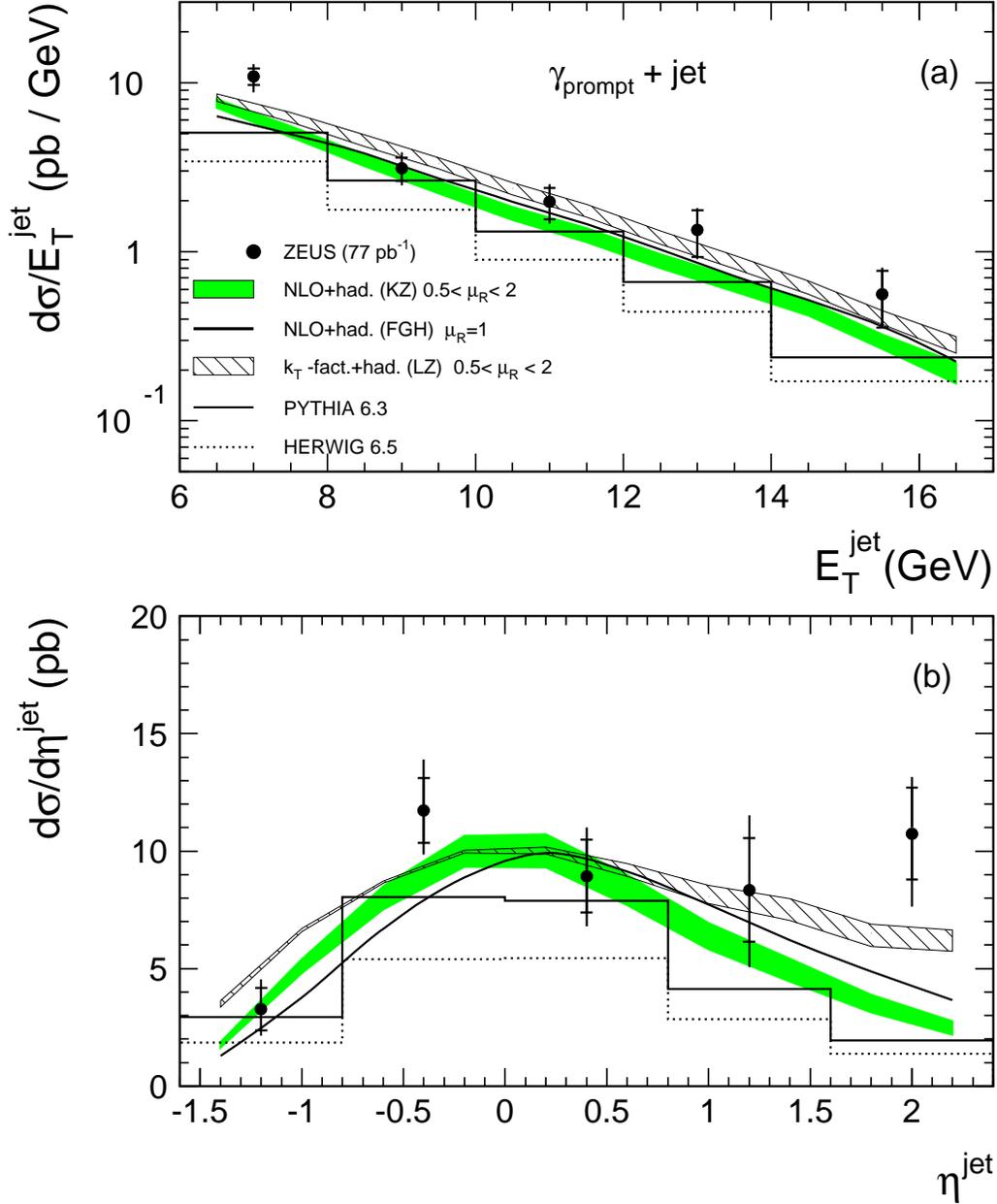}%
\caption{The $\gamma+\mathrm{jet}$ differential cross sections
as functions of $E_{T}^{\rm jet}$ and $\eta^{\rm jet}$ 
compared to the QCD calculations (with hadronisation corrections) 
and Monte Carlo models.}
\label{prph_4}
\end{center}
\end{figure}

\newpage
\begin{figure}
\begin{center}
\includegraphics[height=15.0cm]{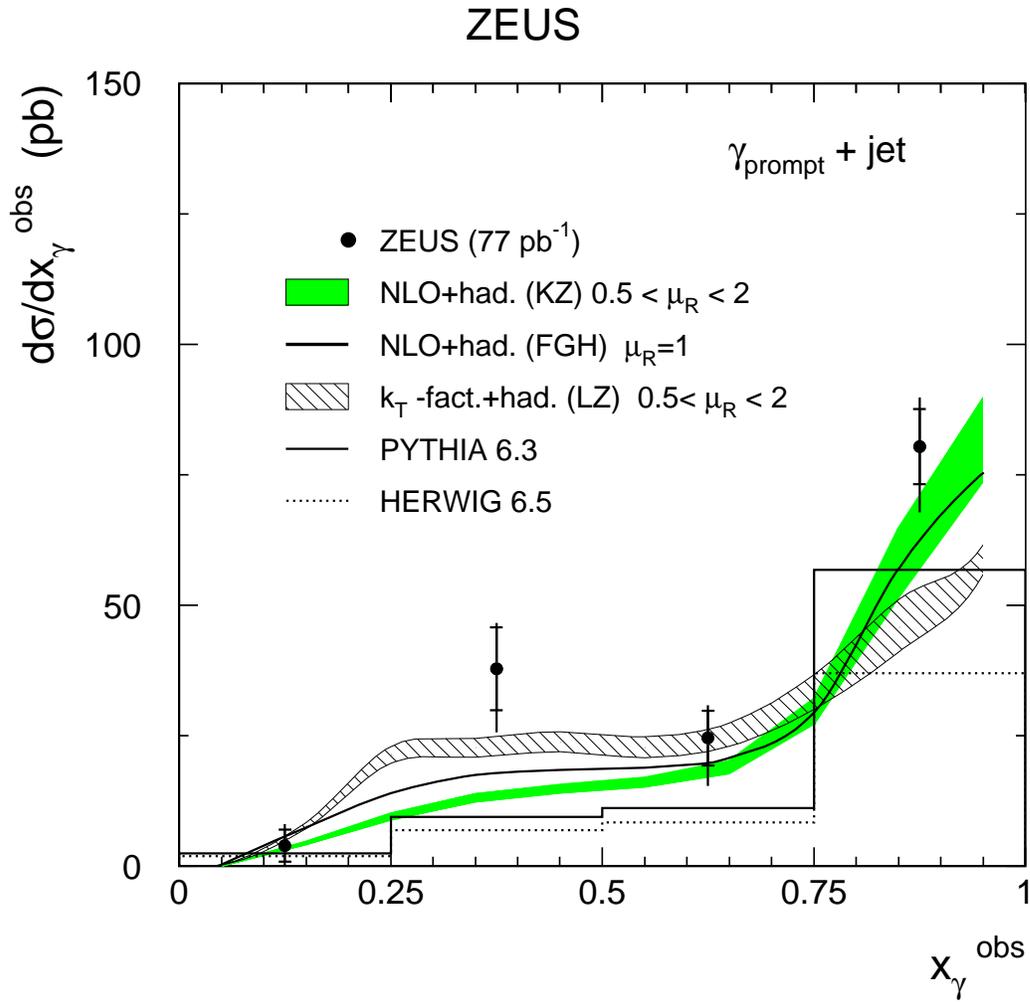}%
\caption{The $x_{\gamma}^{\mathrm{obs}}$  cross section for $\gamma+\mathrm{jet}$ events
compared to the NLO QCD calculations (with hadronisation corrections)
and Monte Carlo models. }
\label{prph_5}
\end{center}
\end{figure}

\newpage
\begin{figure}
\begin{center}
\includegraphics[height=20.0cm]{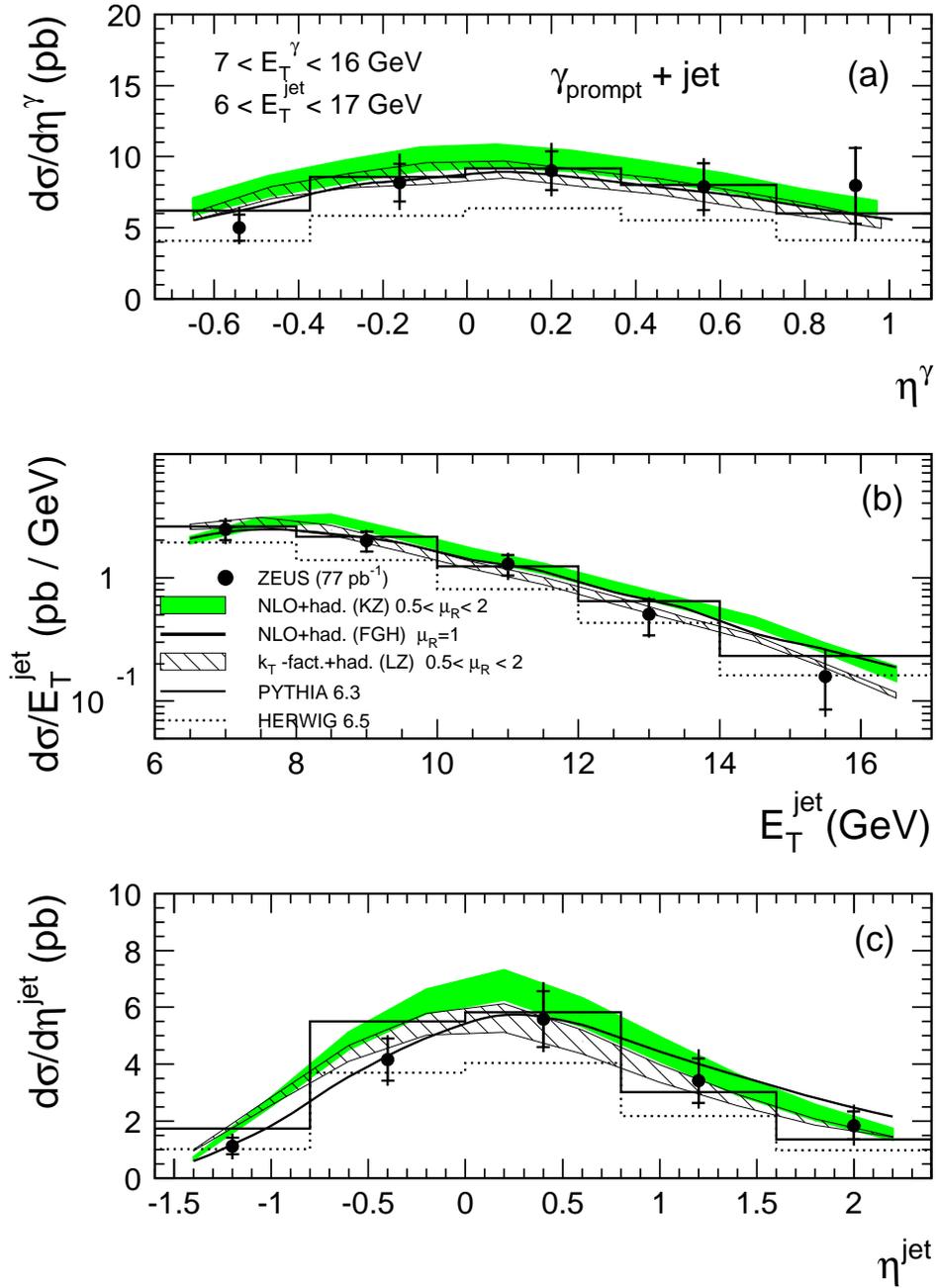}%
\caption{The differential $\gamma+\mathrm{jet}$ cross sections
as functions of: 
a)  $\eta^{\gamma}$,  b) $E_{T}^{\rm jet}$ and c) $\eta^{\rm jet}$ 
compared to the QCD calculations (with hadronisation corrections)
and Monte Carlo models. 
The cuts are the same as for Figs.~\ref{prph_3} and ~\ref{prph_4},
except for the cut on the 
transverse energy of the prompt photons,  which was increased to $7\gev$. 
}
\label{prph_6}
\end{center}
\end{figure}

\newpage
\begin{figure}
\begin{center}
\includegraphics[height=15.0cm]{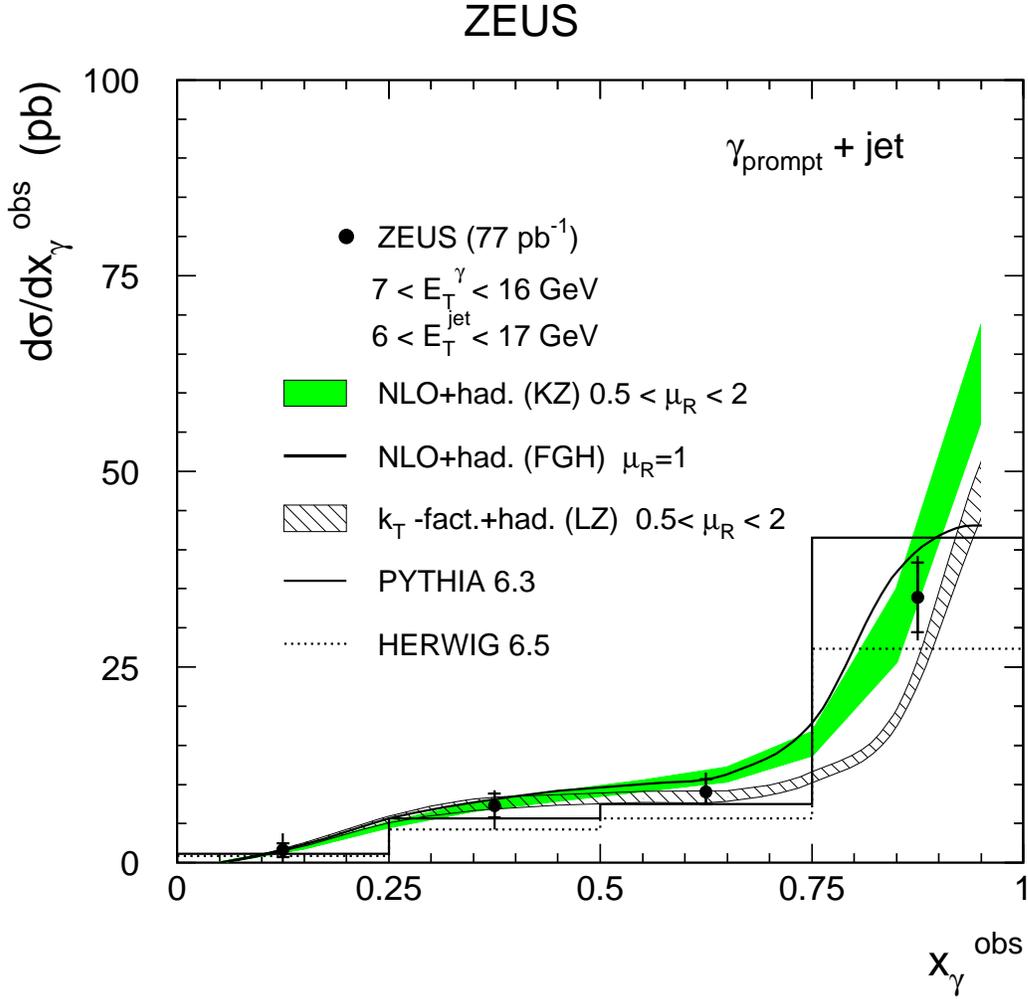}%
\caption{The $x_{\gamma}^{\mathrm{obs}}$  cross section for $\gamma+\mathrm{jet}$ events
compared to the QCD calculations (with hadronisation corrections)
and Monte Carlo models. The cuts are the same as for Fig.~\ref{prph_5}
except for the cut on the
transverse energy of the prompt photon which was increased to $7\gev$.
}
\label{prph_7}
\end{center}
\end{figure}